\documentclass[a4paper]{ar-1col}

\usepackage{enumerate}
\usepackage{todonotes}
\usepackage{graphicx}
\usepackage{floatrow}
\usepackage{amsmath}
\usepackage{amsfonts}
\usepackage{bm}
\usepackage[round,sort&compress]{natbib}
\usepackage{subfig}
\usepackage{xcolor}

\definecolor{lightgreen}{rgb}{.90,1,0.90}

\graphicspath{ {./figs/}}

\jname{Annual Review of Fluid Mechanics}
\fstpage{357}
\endpage{77}
\jyear{2019}
\jvol{51}
\doi{10.1146/annurev-fluid-010518-040547}

\begin{document}

\markboth{Duraisamy, Iaccarino, Xiao}{Turbulence Modeling in the Age of Data}

\title{Turbulence Modeling in the Age of Data}

\author{Karthik Duraisamy$^{1, \star}$, Gianluca Iaccarino$^{2, \star}$, and Heng Xiao$^{3, \star}$
\affil{$^1$Department of Aerospace Engineering, University of Michigan, Ann Arbor, MI 48109; kdur@umich.edu}
\affil{$^2$Department of Mechanical Engineering, Stanford University, Stanford, CA 94305; jops@stanford.edu}
\affil{$^3$Kevin T. Crofton Department of Aerospace and Ocean Engineering, Virginia Tech, Blacksburg, VA 24060; hengxiao@vt.edu}
\affil{$^\star $ These authors contributed equally to this article and are listed alphabetically}}

\begin{abstract}
Data from experiments and direct simulations of turbulence have historically been used to {\em calibrate}  simple engineering models such as those based on the Reynolds-averaged Navier--Stokes (RANS) equations. In the past few years, with the availability of large and diverse datasets, researchers have begun to explore methods to {\em systematically inform} turbulence models with data,
with the goal of quantifying and reducing model uncertainties. This review surveys recent developments in bounding uncertainties in  RANS models via physical constraints, in adopting statistical inference to characterize model coefficients and estimate discrepancy, and in using machine learning to improve turbulence models. Key principles, achievements and challenges are discussed. A central perspective advocated in this review is that by exploiting foundational knowledge in turbulence modeling and physical constraints, data-driven approaches can yield useful predictive models.
\end{abstract}

\begin{keywords}
turbulence modeling, statistical inference, machine learning, data-driven modeling, uncertainty quantification 
\end{keywords}
\maketitle

\section{Introduction}

Turbulence is a common physical characteristic of fluid flows. In wind turbine design, the knowledge of the turbulence 
 in the incoming flow and in the blade boundary layers is important for performance; in internal combustion engines, vigorous 
turbulence increases fuel/air mixing, improving overall efficiency and reducing emissions; in airplane design, delaying the 
occurrence of turbulence in  boundary layers over the wing surfaces leads to reduced  fuel consumption. 
These examples, and a vast number of other applications, demonstrate the importance of determining the effect of turbulence 
on the performance of engineering devices, and justify the continuous interest in developing techniques to simulate and 
predict turbulent flows.

The representation of turbulent motions is challenging because of the broad range of active spatial and temporal scales involved
and the strong chaotic nature of the phenomenon. Over the past half century, starting from the pioneering theoretical 
studies of Prandtl, Kolmogorov, and von Karman,~\citep{darrigol2005worlds} many theoretical and computational approaches have been introduced to characterize turbulence. The continuous growth of 
 computer power has enabled  direct numerical simulations (DNS) of a number of turbulent flows and processes involving the physics of turbulence~\citep{kim1987turbulence,rastegari2018common}.  However, simplified engineering approximations continue to remain 
 popular and widespread across different industries.
Among these, RANS (Reynolds Averaged Navier-Stokes) and
LES (Large Eddy Simulation) approaches are the most common, although there exist many alternatives~\citep{spalart2009detached-eddy,girimaji2006partially-averaged}. RANS techniques rely completely on modeling assumptions to 
represent turbulent characteristics and, therefore, lead to considerably lower computational requirements than DNS.  
RANS models are constructed using a formal averaging procedure
applied to the exact governing equations of motion and require {\it closures} to represent the turbulent stresses and scalar 
fluxes emerging from the averaging process.  The discipline of turbulence modeling has evolved using a combination of intuition,
asymptotic theories and empiricism, while constrained by practical needs such as numerical
stability and computational efficiency.
Single-point RANS models of turbulence, that are the focus of this review, are by far the most popular methods. These 
 models implicitly assume an equilibrium  spectrum and locally-defined constitutive relationships to {\it close} the averaged 
governing equations and express unclosed terms as a function of averaged, local flow quantities.

\begin{marginnote}[]
{The title of this article is inspired by the recent book of \citet{efron2016computer}. It reflects the belief that recent advances in data sciences are 
offering new perspectives to the classical field of turbulence modeling.}
\end{marginnote}

LES techniques, on the other hand, directly represent  a portion of the active scales and only require modeling  to account for 
the unresolved turbulent motions. LES is gaining popularity in many industrial applications characterized 
 by relatively small Reynolds numbers. As LES also involves modeling assumptions, some of the ideas outlined in 
this manuscript regarding RANS are amenable for use in LES~\citep{jofre2018framework, gamahara2017searching}.

The inherent assumptions in the RANS approach and the process of formulating closure models introduce potential 
accuracy limitations and, consequently, reduced credibility in  its predictive ability. Direct quantification of the errors 
introduced by closure models is intractable in general, but formal uncertainty quantification techniques have recently
enabled  the interpretation of RANS predictions in probabilistic terms, while characterizing the corresponding confidence levels.
Experimental observations have routinely been used  to {\it calibrate} the closures and attempt to improve the 
accuracy of the resulting computations.  Statistical inference approaches  enable a more comprehensive {\it fusion}
of data and models, resulting in improved predictions. Furthermore, the introduction of modern machine learning strategies
brings fresh perspectives to the classic problem of turbulence modeling. 

\begin{textbox}[h]
\section{Lexicon of data-driven modeling}
The elements of a data-driven model are:
\begin{itemize}
\item[$\mathcal{M}$:] the (computational) {\it model}, which is typically a function of an array of independent variables $\mathbf{w}$ and based on a set of algebraic or differential operators $\mathcal{P}$; the model  includes a set of parameters $\mathbf{c}$ that are the {\it primary} target of the data modeling;
\item[$\bm{\theta}$:]  the {\it data}, which in general is accompanied by a quality estimate, or in other words, the  uncertainty $\bm{\epsilon}_{\bm{\theta}}$;
\item[$\bm{o}$:] the {\it output} of the model corresponding to the data $\bm{\theta}$; 
\item[$\bm{\delta}$:] the {\it discrepancy} which describes the ability of the model to {\it represent} the data. In general, $\bm{\delta}$ is a function of the model and it is unknown; it is typically described in terms of $\bm{\theta}$ and a set of {\it features} $\bm{\eta}$ that are derived from prior knowledge, constraints or directly from data. 
\end{itemize}
A general data driven model is written as
\[
\widetilde{ \mathcal M} \equiv \mathcal M ( \mathbf{w};  \mathcal P(  \mathbf{w}  );  \mathbf{c}; \bm{\theta}; \bm{\delta};  \bm{\epsilon}_{\bm{\theta}} )
\]
and one is generally interested in predicting quantities of interest $\bm{q}(\widetilde{\mathcal M})$.

\end{textbox}

\section{Turbulence closures  and uncertainties}
\label{sec:levels}
Assumptions are introduced in several stages 
whilst constructing a  Reynolds-averaged model. Although fully justified under certain conditions, these 
hypotheses introduce potential inadequacies that limit the credibility of the overall predictions if not
properly quantified. In this section, we illustrate the  four layers of simplifications that are typically required to 
formulate a RANS closure.

\begin{itemize}

\item[L1:]
The application of  time- or ensemble-averaging operators $\langle \cdot \rangle$ combined with the non-linearity of the Navier-Stokes equations (indicated hereafter as $\mathcal N(\cdot) = 0$)
leads to an undetermined system of equations, that requires the introduction of modeling assumptions to {\it close} the
system.  
\begin{marginnote}[]
\entry {L1} {Uncertainties introduced by ensemble-averaging, which are fundamentally irrecoverable.}
\end{marginnote}
\begin{equation}
\langle \mathcal N(\cdot) \rangle \ne \mathcal N( \langle \cdot \rangle )
\label{eq:L1a}
\end{equation}

At a given instant in time, there are infinite realizations of velocity fields (microscopic state)
that are compatible with an averaged field (macroscopic state); however each of these realizations might  evolve dynamically in different ways, leading
to hysteresis-like phenomena and thus uncertainty. This  inadequacy is unavoidable in RANS, due to the loss of information 
in the  averaging process and is fundamentally irrecoverable regardless of the sophistication of the turbulence model. This situation is not limited to turbulence modeling --  and in general terms, is referred to as {\it upscaling} or {\it coarse graining}~\citep{rudd1998coarse-grained}.

 \begin{marginnote}[]
\entry {L2} {Uncertainties in the functional and operational representation of Reynolds stress}
\end{marginnote}

\begin{itemize}
\item[L2:]
In the process of developing closures, a model representation is invoked to relate the macroscopic state to the microscopic state
and  formally remove the  unknowns resulting from the averaging process.  

\begin{equation}
\langle \mathcal N(\cdot) \rangle =  \mathcal N( \langle \cdot \rangle ) + \mathcal M (  \cdot  ).
\label{eq:L2}
\end{equation}

For an incompressible fluid, the unclosed term is simply written as:

\begin{equation}
 \mathcal M (  \cdot  ) =  {\nabla} \cdot  \boldsymbol{\tau},
 \label{eq:L1b}
\end{equation}

 \noindent where $\boldsymbol{\tau}$ is the Reynolds stress tensor.

$\mathcal{M}$ is  written in
terms of a set of independent, averaged variables $  \mathbf{w} $,  defined either locally or globally, leading to one-point or two-point closures. 
For instance, with the assumption that the Reynolds stress tensor is only
a function of the local, averaged velocity gradient tensor, the Cayley-Hamilton theorem can be applied to derive an exact expansion basis~\citep{gatski2000nonlinear}. Linear eddy viscosity models and algebraic stress models are examples of L2-level assumptions.

\begin{marginnote}[]
\entry {L3} {Uncertainties in functional forms within a  model}
\end{marginnote}

\begin{itemize}
\item[L3:]
Once the independent variables are selected, a specific functional form is postulated. Either algebraic or differential equations, denoted here as $\mathcal P (  \cdot  )$, are typically used to 
 represent physical processes or specific assumptions. Schematically, the model is now:
 \begin{equation}
 \mathcal M (   \mathbf{w}; \mathcal P(  \mathbf{w}  ) ).
 \label{eq:L3}
 \end{equation}
  One and two-equation models are the most popular, although
 many different choices of the independent variables exist in  literature~\citep{wilcox2006turbulence}. Often $\mathcal P(  \cdot  )$ mimic the terms in the
 Navier-Stokes equations such as convection and diffusion; additional contributions and source terms are often included to represent known sensitivities,
 such as near wall dynamics, rotational  corrections, etc.~\citep{durbin2017some}.  Although it is possible to derive differential models formally through repeated applications of the Navier-Stokes operator ($  \mathcal N(  \boldsymbol{\tau}  ) = \cdot $),  this leads to computationally intensive closures and numerically cumbersome and non-interpretable terms.  

\begin{marginnote}[]
\entry {L4} {Uncertainties in the coefficients within a model}
\end{marginnote}

\begin{itemize}
\item[L4:]
 Finally, given a complete model structure and functional form, a set of coefficients $\mathbf{c}$ must be specified to calibrate the relative importance of the various
 contributions in the closure. Formally the closure is then:
 \begin{equation}
 \mathcal M (   \mathbf{w}; \mathcal P(  \mathbf{w}  );  \mathbf{c} ).     
  \label{eq:L4}
 \end{equation}

 It is common to use consistency between the closure and known asymptotic turbulence states to define some of the coefficients,
 although in many cases empirical evidence is  more effective in producing realistic models~\citep{durbin2017some}. The choice of the $C_\mu$ coefficient in two-equation linear eddy viscosity models is a classical L4 closure issue.
 \end{itemize}
 \end{itemize}
 \end{itemize}
 \end{itemize}
 
A RANS prediction of a quantity of interest  $ \mathbf{q}$ is then in general
\begin{equation}
 \mathbf{q} = \mathbf{q} \left(   \mathcal N( \langle \cdot \rangle ); \mathcal M (   \mathbf{w}; \mathcal P(  \mathbf{w}  );  \mathbf{c} ) \right)
  \end{equation}

Together, these four modeling layers showcase the difficulty in assessing the true predictive nature of a turbulence closure, the inconsistency inherent in comparing different strategies, and the need for a
careful and transparent process for quantifying model inadequacies and reducing them.

\begin{textbox}[h]
\section{Uncertainty Quantification}

In spite of the considerable popularity of simulations in science and engineering, the process of generating objective confidence levels 
in numerical predictions remains a challenge. The complexity arises from (a) the imprecision or natural variability in the inputs to any 
simulation of a real-world system ({\it aleatory} uncertainties), and (b) the limitations intrinsic in the
physics models ({\it epistemic} and {\it model-form} uncertainties). 
Uncertainty Quantification (UQ) aims to  rigorously measure and rank  the effect of these uncertainties on prediction outputs. 

The first step in UQ is  the identification of the sources of uncertainty and the introduction of an appropriate description (typically in probabilistic terms) $\bm{\epsilon}$;
this is then {\it propagated} through the  model $\mathcal{M}$, resulting in predictions of a quantity of interest $\mathbf{q}( \mathcal{M}, \bm{\epsilon})$.
The propagation step is typically computationally intensive and has received considerable attention in the last decade, leading to the development of extremely efficient 
UQ strategies. The result is a prediction that  explicitly represents the impact of the uncertainty; if $\bm{\epsilon}$ is a stochastic quantity, the resulting
prediction is the probability distribution $ \mathbb P(\mathbf{q})$ and, therefore, a rigorous measure of the confidence interval can be extracted from the analysis.
In some cases, only statistical moments of $\mathbf{q}$ are required, leading to  more cost-effective UQ propagation strategies.
Alternative descriptions of the uncertainties are also possible. For instance, when very limited observations are available to represent a specific uncertainty source,
it is appropriate to  introduce a range (an {\it interval} $[\bm{\epsilon}^-;\bm{\epsilon}^+]$)  and consequently seek an interval on the model predictions  $[\mathbf{q}^-; \mathbf{q}^+]$.
In general, for non-linear models $\mathbf{q}^{\pm} \ne \mathbf{q}( \mathcal{M}, \bm{\epsilon}^{\pm} )$. In such cases, optimization techniques and bounding strategies are used instead
of probabilistic approaches.

\end{textbox}

\section{Models, data and calibration}
\label{sec:models-data}

The use of experimental observations to  drive physical insight is a staple of the scientific method.  The understanding of turbulent flows has benefited  
considerably from detailed measurement campaigns such as the famous  isotropic turbulence experiments of \citet{comte1966}. The measured 
turbulence decay rates have been used to constrain the value of the (L4) constants $\mathbf{c}$ defined earlier.

In the last four decades, in addition to experimental datasets, direct simulations of the Navier-Stokes equations (DNS) have provided a further, invaluable source of
data to gather modeling insights. The Summer Program of the Center for Turbulence Research at Stanford University had been established with the  goal of
\emph{studying turbulence using numerical simulation databases} in 1987. There has been a concerted effort by the turbulence community to gather and archive   data sets including the databases in US \citep{l12008} and Europe~\citep{coupland1993}, among others. Until the past decade however,  experimental and simulation data have been used mostly to obtain modeling insight and 
to aid in validation. Recently, data has been used towards the end of systematically informing turbulence models with the goal of quantifying and reducing model uncertainties.

In this section we describe how data   is used in building  new, calibrated models $\widetilde{ \mathcal M} \ne { \mathcal M}$.

\subsection{Naive calibration}
 
The simplest calibration process typically involves the  selection of an experimental configuration that is similar to that of 
the prediction target. Often, the measured data  may be the same as the quantity of interest $ \mathbf{q}$, but available 
  in different flow conditions or configurations. Uncertainties in the measurements are typically ignored. Finally it is assumed that the model coefficients ($\mathbf{c}$)
 are the dominant source of uncertainty in the model and, therefore, the calibrated model is:
 
  \begin{equation}
\widetilde{ \mathcal M} = \mathcal M (   \mathbf{w}; \mathcal P(  \mathbf{w}  );  \tilde{ \mathbf{c}}_\mathbf{q}),    
 \end{equation}

\noindent and the prediction accuracy is judged by the difference in $\mathbf{q}$ obtained when using $\mathbf{c}$ or $ \tilde{ \mathbf{c}}_\mathbf{q}$.
 This process has led to the proliferation of turbulence model variants and the inherent difficulty in assessing predictive capabilities.

\subsection{Statistical inference}

Statistical inference is the generalization of the calibration process described above; specifically, uncertainty in the experiments can be directly accounted for and
a potential  discrepancy (misfit) between the model prediction $ \bm{\delta}$ and the data is also included. Furthermore, the calibration data can include evidence from different
sources while the  objective is simply to represent the data. The inference is  formulated in a probabilistic setting, inspired by the Bayes theorem and the result is a calibrated, stochastic model:

   \begin{equation}
 \widetilde{ \mathcal M} = \mathcal M (   \mathbf{w}; \mathcal P(  \mathbf{w}  );    \tilde{ \mathbf{c}}_{\bm{\theta}} ) +  \bm{\delta} + \bm{\epsilon}_{\bm{\theta}}.    
 \label{eq:inteference} 
 \end{equation}

Formally, stochasticity is a consequence of uncertainty in the measurements, the prior information on the calibration parameters (for example, the range or the most likely values of $\mathbf{c}$), and the discrepancy function. A prior for the discrepancy function is typically left to the intuition of the modeler and is typically represented in a simple mathematical form, for example using a Gaussian random field with parameters that are also estimated through the calibration process, i.e. $\bm{\delta}(\bm{\theta})$. The inference problem is typically solved using Markov-Chain Monte Carlo (MCMC) strategies, and the result is a stochastic description (the posterior probability) of the model $ \widetilde{ \mathcal M}$.

The resulting model prediction $\mathbf{q}( \widetilde{ \mathcal M} )$ is  a random quantity.
This approach is referred to in the literature 
as a Bayesian inversion strategy.  
  In many cases,  only the {\it mode} of the the posterior probability is used; this is referred to as the maximum a posteriori (MAP) estimate and therefore the corresponding calibrated model is deterministic.

\begin{textbox}[h]
\section{Statistical inversion }

Statistical inversion aims to identify parameters $\bm{c}$ of a model ${\mathcal M (\bm{c} )}$ given data $\bm{\theta}$  with  uncertainty $\bm{\epsilon}_{\bm{\theta}}$. Mathematically, the solution 
is determined by minimizing the difference between $\bm{\theta}$ and the corresponding model output $\bm{o}({\mathcal M}(\bm{c}))$.
In a Bayesian framework, the result of the inversion, i.e. the {\em posterior} probability of $\bm{c}$ given $\bm{\theta}$, is given as
\[
\mathbb{P}[ \bm{c} | \bm{\theta}]  \propto  \mathbb{P}[\bm{\theta} |  \bm{c}] \times \mathbb{P}[  \bm{c}],
\]

\noindent where $\mathbb{P}[ \bm{c}]$ is the  {\em prior}, i.e.  the probability of the model before any data (evidence) is used, and   $\mathbb{P}[\bm{\theta} |  \bm{c}]$ is the probability of the model being consistent with the data, referred to as  the \emph{likelihood}.

The prior and the posterior represent the modeler's belief on the probability of $\mathcal{M}$, before and after  observing the data.
In the case that all underlying probability distributions are assumed to be Gaussian, it can be shown \citep{aster2011parameter} that the \textit{maximum a posteriori} (MAP) estimate of $\bm c$ can be determined by solving a deterministic optimization problem
\[
\bm{c}_\textrm{MAP} = \operatorname{arg} \text{min } \frac{1}{2}\bigg[\big(\bm{\theta} - \bm{o}(\mathcal{M}(\bm{c}))\big)^T { \bm{Q}^{-1}_{\bm{\theta}}}\big(\bm{\theta} - \bm{o}(\mathcal{M}(\bm{c}))\big) + \big(\bm{c} - \bm{c}_{prior}\big)^T \bm{Q}^{-1}_{\bm{c}}\big(\bm{c} - \bm{c}_{prior}\big)\bigg],
\]
where ${ \bm{Q}_{\bm{\theta}}}$ and ${ \bm{Q}_{\bm{c}}}$ are the observation and prior covariance matrices, respectively. 

An alternative strategy  to solve inverse problems is the Least Squares (LS) approach. The LS procedure  
involves the minimization of the discrepancy between $\bm{\theta}$ and the model output $\bm{o}(\mathcal{M}(\bm{c}))$ by solving the optimization problem 

\[
\bm{c}_\textrm{LS}  = \operatorname{arg} \text{min } || \bm{\theta}-\bm{o}(\mathcal{M}(\bm{c})) ||_2^2 + \gamma ||\bm{c}||_2,
\]
\noindent where the second contribution scaled by $\gamma$ is a regularization term included to improve well-posedness and conditioning of the inversion process  ($\gamma$ is a user-specified parameter).
Again, assuming that all distributions are Gaussian (and $\gamma \equiv \bm{Q}_{\bm{\theta}} \bm{Q}_{\bm{c}}^{-1}$), $\bm{c}_\textrm{LS} = \bm{c}_\textrm{MAP}$.

\end{textbox}

\subsection{Data--driven modeling}

In the last two decades, the introduction of computationally efficient statistical inference algorithms has led to the possibility of assimilating large amounts of data (for example, those generated by DNS simulations). This   has spurred  interest in approaches that rely more on the available data than on traditional  models; in other words 
in  Eq. \ref{eq:inteference}, the emphasis is on $ \bm{\delta}$ rather than on $\mathcal M$. Different choices for the functional representation of $ \bm{\delta}$ are
available, with increasing focus on  machine-learning strategies. 
In addition to the inference, further work has been devoted towards representing the
  discrepancy $ \bm{\delta}$ in terms of  features $\bm{\eta}$ selected from  a potentially large set of candidates. This enables representation of the resulting model 
  in terms of quantities, such as the mean velocity gradients, that are likely to be {\it descriptive} in a more general context than the one characterized by the available data.
  Furthermore,  constrains such as symmetry properties or Galilean invariance  can be enforced in the definition of the candidate features.  
  
In general, data--driven models can then be expressed as:

  \begin{equation}
 \widetilde{ \mathcal M} = \mathcal M (   \mathbf{w}; \mathcal P(  \mathbf{w}  );    \mathbf{c}(\bm{\theta}); \bm{\delta}{(\bm{\theta},\bm{\eta})};   \bm{\epsilon}_{\bm{\theta}}  ).   
 \label{eq:datadriven} 
 \end{equation}

\subsection{Calibration and prediction}

The objective of the calibration process is the definition of a model that incorporates evidence from available data. In practical applications, the 
next step is to use  the calibrated model to predict a quantity of interest. 
In the seminal work of~\citet{kennedy2001bayesian}, model-form uncertainty is introduced to the
 prediction by adding a discrepancy term to the model output $\bm{o}(\mathcal{M}(\bm{c}))$. 
Typically, Gaussian Process models are assumed for the model discrepancy $\bm{\delta}$, and Bayesian inversion is used to
derive a posterior distribution for the hyperparameters of the Gaussian process as well as the inferred model parameters.
However, in this approach, the the entire mapping between the input and  prediction is treated as a black-box and thus physics-agnostic.
In the present review, we focus on methods that embed the calibration inside the model.
The propagation of the relevant stochastic information that defines the calibrated model endows the prediction with 
uncertainty  that represents the ability of the calibrated model to represent the data.

\section{Quantifying uncertainties in RANS models}

Predictions based on RANS models are   affected by the assumptions invoked in the construction of the closure (L1--L4) and by the calibration process. A
rigorous   process is required to characterize the potential impact of these sources of uncertainties and the resulting confidence in the predictions.
In this section, we review approaches that seek to derive $ \widetilde{ \mathcal M} = \mathcal M + \bm{\epsilon}_{ \mathcal M}$
where $\bm{\epsilon}_{ \mathcal M}$ is either obtained from theoretical arguments or $\bm{\epsilon}_{ \mathcal M} = \bm{\epsilon}(\bm{\theta})$ from comparisons to existing data.

\subsection{Uncertainties in the Reynolds stress tensor}

We survey two strategies to characterize the uncertainty in the Reynolds stresses based on an interval and a probabilistic description of the uncertainty, respectively.

RANS modeling has traditionally aimed to approximate unclosed terms in the averaged equations with the goal of obtaining
 a {\it closed}, computable  model. An alternative idea is to replace the unclosed terms with {\it bounds} that are 
 based on theoretical arguments, leading to predictions that represent extreme but possible behaviors, rather than likely behaviors under specific assumptions. In other words, bounding aims to construct prediction intervals that 
 can be proven to {\it contain} the true answers, as opposed to 
  explicit estimates that might be inaccurate. In the 70s, pioneering work in turbulence analysis by~\citet{howard1972bounds},~\citet{busse1970bounds} and others, and more recently the work by Doering and coworkers~\citep{doering1994variational},  explored this idea;
 these authors applied variational approaches and formulated a generic framework for estimating bounds on
 physical quantities rigorously and directly from the equations of motion. Following this approach, named the {\it background flow method},
 one manipulates the equations of motion relative to a steady trial background state and
 then decomposes the quantity of interest, for example the energy dissipation rate, into a background profile
 and a fluctuating component. If the fluctuation term satisfies a non-negativity condition, the background part admits an upper bound.
 A recent improvement of this approach \citep{seis2015scaling} led to quantified bounds on the  energy dissipation rate for 
 shear flow, channel flow, Rayleigh--Benard convection and porous medium convection.
 In spite of these encouraging results, it is difficult to envision that formal bounds can be derived for  flow problems of practical
 engineering interest.

\begin{figure}[!htbp]
  \centering
   \begin{floatrow}
    \subfloat[Turbulence componentality in Barycentric triangle ~\citep{jofre2018framework}] {\includegraphics[width=0.43\textwidth]{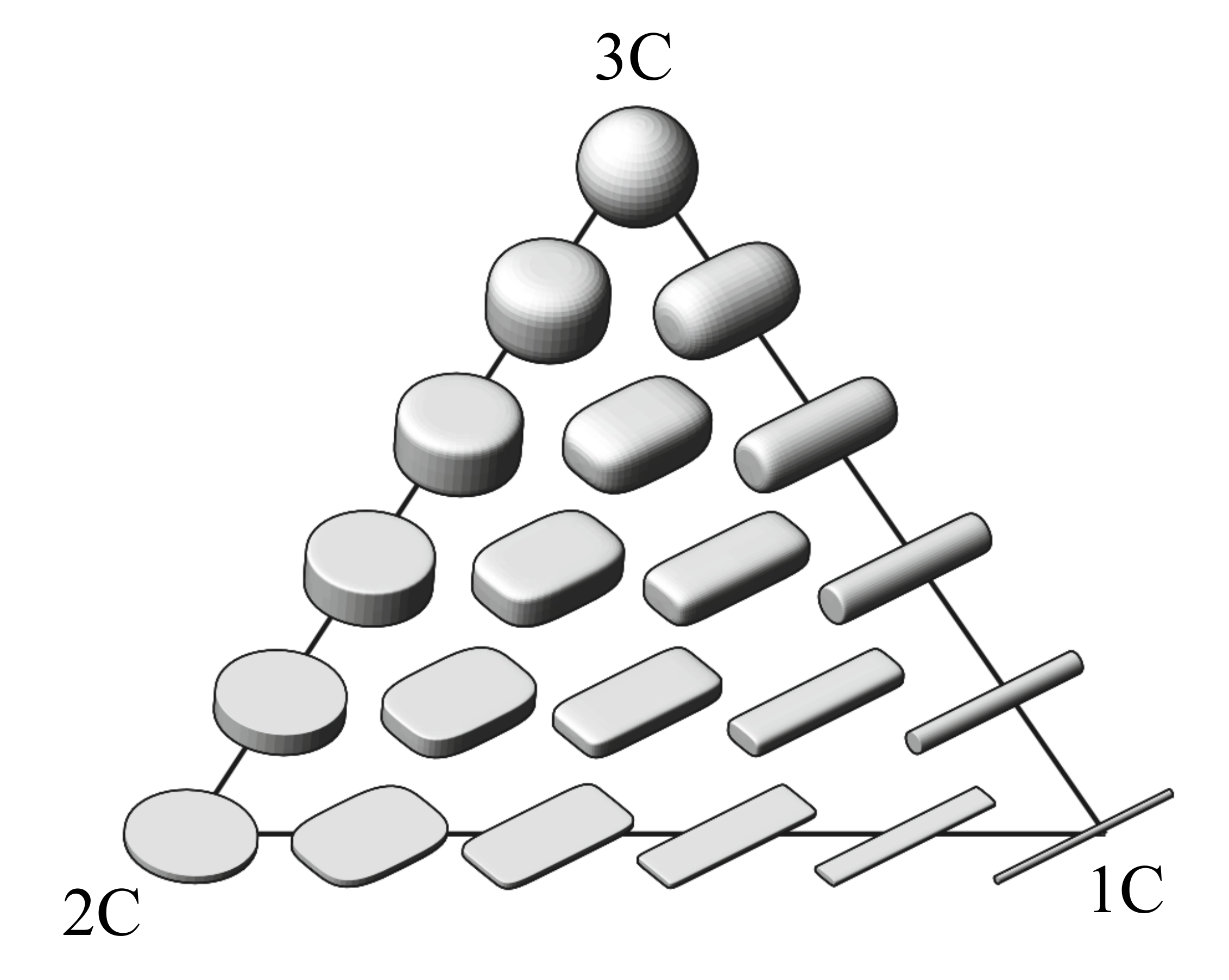}}  
  \subfloat[Realizability-constrained perturbations] {\includegraphics[width=0.52\textwidth]{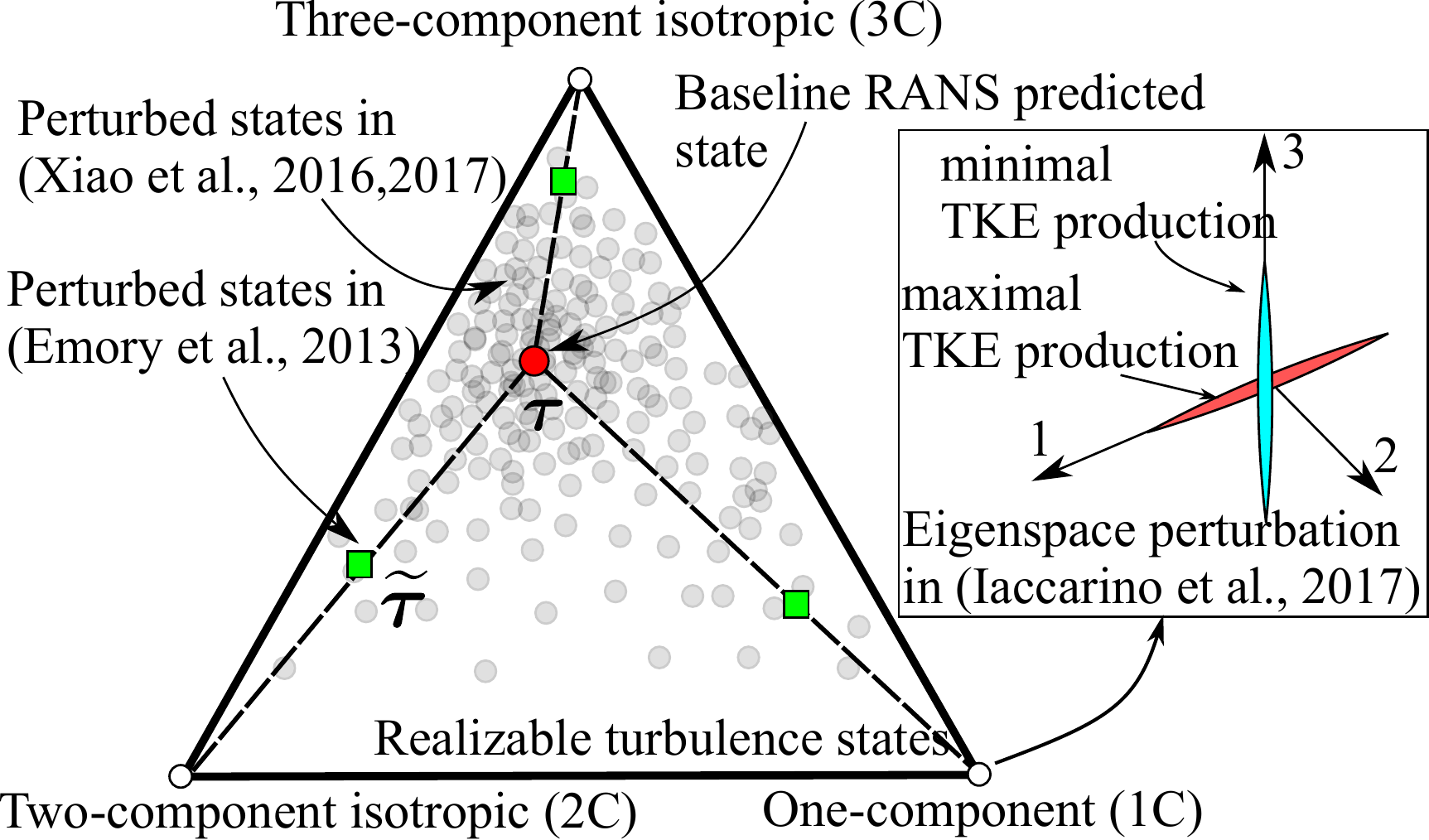}}
 \end{floatrow}
 \vspace{1em}
  \caption{Geometric representation of the realizability constraint on a single-point Reynolds stress tensor.}
  \label{fig:realizable}
\end{figure}

An alternative viewpoint on bounding is based on the concept of Reynolds stress realizability \citep{schumann1977realizability, lumley1978computational, pope1985pdf, durbin1994realizability}.
 \cite{emory2011modeling,emory2013modeling} proposed a scheme of introducing 
realizable, physics-constrained perturbations to the
Reynolds stress tensor as a way of quantifying (L2) uncertainties in RANS models expressed as intervals. 
The starting point is the eigen-decomposition of the Reynolds stresses. \citet{banerjee2007presentation} defined a mapping based on  the eigenvalues $\lambda_1$, $\lambda_2$, and $\lambda_3$ and
corresponding  Barycentric coordinates that enables a convenient visual representation of the realizability bounds in a two-dimensional coordinate system,  depicted in Fig.~\ref{fig:realizable}.

Formally a general expression for the Reynolds stresses can be written as: 
\begin{equation}
  \widetilde{\boldsymbol{\tau}} = \boldsymbol{\tau}^{\text{RANS}} +  \boldsymbol{\delta}_{\bm{\tau}}
  = 2 \widetilde{k} \left( \frac{1}{3} \mathbf{I} + \widetilde{\mathbf{V}} \widetilde{\bm{\Lambda}} \widetilde{\mathbf{V}}^{\top} \right).
  \label{eq:eigen}
\end{equation}
where $\widetilde{k}$, $\widetilde{\bm{\Lambda}}$, and $\widetilde{\bm{V}}$ are the perturbed counterparts to Reynolds stresses computed using a RANS model.  For example, for the eigenvalues $\widetilde{\bm{\Lambda}} = \bm{\Lambda}^{\text{RANS}} + \boldsymbol{\delta}_{\lambda}$.

\begin{figure*}
\centering
\includegraphics[width=0.92\textwidth]{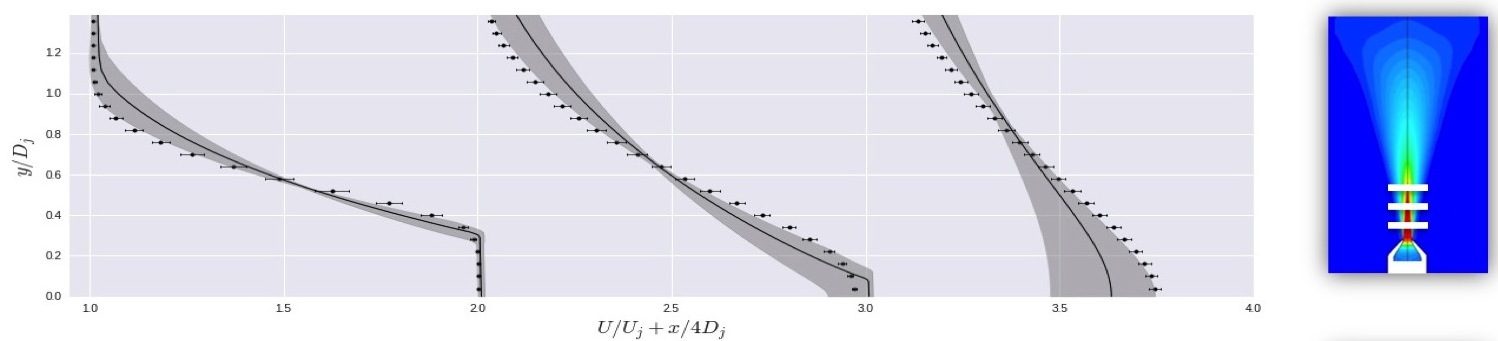}
\caption{Prediction bounds for the velocity profiles in a turbulent jet computed using the eigenspace perturbations \citep{mishra2017jets}. The line represent the prediction using an eddy viscosity model, the symbols are the experimental measurements and the grey areas are the computed uncertainty bounds.}.
\label{fig:jet_bounds}
\end{figure*}

With the objective of determining general bounds, 
 the discrepancy $ \boldsymbol{\delta}_{\bm{\tau}} = (\boldsymbol{\delta}_{\lambda},\boldsymbol{\delta}_k,\boldsymbol{\delta}_\mathbf{V} )$  has to be defined with  appropriate physical constraints and without the use of calibration data.
The reliazability condition of the Reynolds stress  depicted in Fig. \ref{fig:realizable} provides clear constraints on the eigenvalues.
In Emory et al.'s approaches, RANS predicted Reynolds stresses were perturbed towards three representative limiting states: one-component (1C), two-component (2C), and three-component (3C) turbulence, indicated by the vertices of the Barycentric triangle. 
In contrast to the strong constraint imposed by the realizability on the eigenvalues, the constraint on the turbulent kinetic energy is rather weak -- it only has to be pointwise non-negative. Furthermore, the realizability condition does not give  clear bounds on the eigenvectors. 
\citet{iaccarino2017eigenspace} used the two limiting states at which the turbulence kinetic energy (TKE) production \(P = \tau_{ij} {\partial U_i}/{\partial x_j}\)
achieves maximum and minimum values. These states  are identified by specific alignments between Reynolds stress tensor $\tau_{ij}$ and mean velocity gradient tensor ${\partial U_i}/{\partial x_j}$. 
A prediction with a resulting interval-based uncertainty is obtained by performing simulations corresponding to the extreme componentality states, and the two set of eigenvectors providing minimum and maximum kinetic energy production. The computed bounds $[\mathbf{q}^-; \mathbf{q}^+ ]$ appear to provide a satisfactory representation on the uncertainty in the model as reported in Fig. \ref{fig:jet_bounds} for a simple turbulent jet.

As an alternative to the physical approach based on eigen-decomposition, \citet{xiao2017random} pursued a probabilistic description of the Reynolds stress uncertainty. They modeled the true  stress as a random matrix $\bm{T}$ defined on the set 
of symmetric positive semi-definite $3 \times 3$ matrices. The expectation of random matrix $\bm{T}$ is specified to be the RANS modeled Reynolds stress $\bm{\tau}^{\text{RANS}}$, i.e.  \(\mathbb{E}[\bm{T}] = \bm{\tau}^{\text{RANS}}\).
They further defined a maximum entropy distribution for the Reynolds stress tensor, which is sampled to indicate the uncertainty of the Reynolds stresses.

The random matrix approach and the eigen-decomposition based approach are similar in the sense that they ensure  realizability when perturbing the Reynolds stress tensor or sampling from the distribution thereof. The random matrix  approach explores the uncertainty space of eigenvalues and eigenvector simultaneously~\citep{wang2016quantification}, with the correlation among them implicitly specified through the maximum entropy distribution defined on random matrix $\bm{T}$; it  lacks, however, the clear interpretations of the limiting states as in the physics-based, eigen-decomposition approach.

Both  approaches have focused on the error bound of the Reynolds stress at a \emph{single point}.
 A potentially important source of uncertainty comes from the spatial variation of the Reynolds stress discrepancy as the \emph{divergence} of the Reynolds stress field  appears in the RANS  equations; in other words, it is likely that $\bm{\delta}=\bm{\delta}(\bm {x})$. 
  \cite{emory2013modeling} and  \citet{gorle2014deviation} specified a spatial field for the eigenvalue perturbations based on the assumed limitations 
   of  RANS models, while  \citet{wang2016incorporating} and \cite{xiao2017random} used a non-stationary Gaussian process to encode the same empirical knowledge.
\citet{edeling2017data-free} proposed a ``return-to-eddy-viscosity model'', which is a transport equation with a source term describing the departure of turbulence state from local equilibrium. 
Finally, \citet{wu2018pde-informed} utilized the fundamental connection between the governing partial differential equations and the covariance  of a discrepancy field to derive a physically consistent covariance structure.
\citet{xiao2018quantifying} provided more detailed discussions on some of the approaches above.

\subsection{Uncertainty in model parameters}

The model parameters in turbulence models are often determined enforcing consistency in the prediction of fundamental flows (e.g., homogeneous isotropic turbulence, logarithmic layer). It is well known that these parameters are not universal and might require flow-specific adjustments. For example, \citet{pope2000turbulent} and \citet{eisfeld2017reynolds}  list  optimal parameters for several typical free shear flows (e.g., plane jet, round jet, wake). However, for the lack of better alternatives, the default parameters determined from the fundamental flows are still used in the simulation of complex turbulent flows: this lead to uncertainties (L4).

It is fairly straightforward to assess the impact of uncertainties in the choice of the coefficients in the models by using classical uncertainty propagation techniques. However,
this exercise is fundamentally dependent on the choices made to describe the parameters, i.e. to define their range or their probabilistic representation. A more effective approach is to use data to infer the parameters and then propagate the resulting stochastic description through the simulations
obtaining predictions with uncertainty intervals. This will be discussed in the next section because it is akin to a data-informed approach in which $  \widetilde{ \mathcal M} = \mathcal M (\bm{\theta}) + \bm{\epsilon}_{ \bm{\theta}}$.

\subsection{Identifying regions of uncertainty}

While the above discussion was focused on {\em quantifying} uncertainties,  techniques have been developed to {\em identify} regions of potentially high uncertainties in RANS predictions.
\citet{gorle2014deviation} developed an analytical marker function to identify regions  that deviate from parallel shear
flow. This marker  was found to correlate well with regions  where the prediction of the Reynolds stress divergence was inaccurate. \citet{ling2015evaluation} employed databases of DNS and RANS solutions and formulated the evaluation of potential adequacy of the RANS model as a \emph{classification} problem in  machine learning. The results include several fields of binary labels that indicate whether the specified model assumption is violated.
Although these studies are useful to illustrate the failure of turbulence models there is no straightforward way to use the results for improving predictions.

 \begin{figure}[!htbp]
  \centering
  \begin{floatrow}\hspace{1.5cm}
 \includegraphics[width=0.77\textwidth]{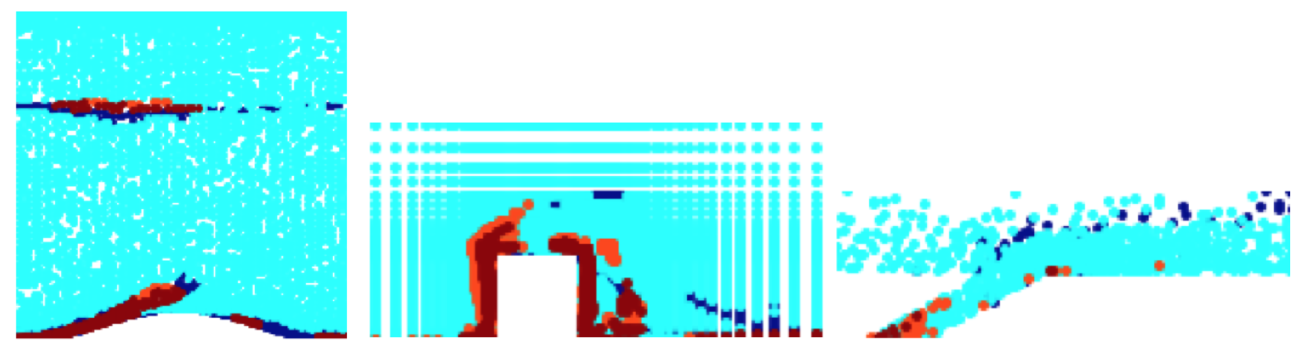} %
  \includegraphics[width=0.18\textwidth]{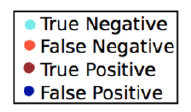} %
 \end{floatrow}
 \vspace{0.5em}
  \caption{Random forest predicted markers (blue and maroon) showing regions with possible emerging of negative eddy viscosity in several flows, indicating likely failure of linear eddy viscosity models in such regions~\citep{ling2015evaluation}. 
  \label{fig:markers}}
\end{figure}

\section{Predictive modeling using data-driven techniques}

The efforts discussed in the previous section aim at providing confidence in the application of RANS closures by identifying and quantifying  uncertainties in the models at various levels. In this section, we focus on approaches that attempts to improve the overall prediction accuracy by using data. In this section, we review strategies that seek to derive $  \widetilde{ \mathcal M} = \mathcal M (\bm{\theta}) $, while explicitly introducing a discrepancy function $\bm{\delta}$.

\subsection{Embedding inference-based discrepancy}

The use of rigorous statistical inference to determine model coefficients is only a relatively recent development.
\citet{oliver2011bayesian} and \citet{cheung2011bayesian} were the first to leverage DNS data from plane channel flows and Bayesian inference  to assign posterior probability distributions to model parameters of several  turbulence models. \citet{edeling2014bayesian,edeling2014predictive} further introduced the concept of Bayesian model scenario averaging in the calibration of model parameters to assess the effectiveness of diverse source of data in specific predictions. Their studies included data from a number of wall-bounded flows.
 \cite{lefantzi2015estimation} and \citet{ray2016bayesian} used a similar approach to infer the RANS model coefficients and also investigated  the likelihood of competing closures while  focusing  on  jet-in-cross-flow, which is a canonical flow in film cooling in turbo-machinery applications. More recently,
\citet{ray2018learning} used experimental data and Bayesian inference to calibrate the parameters in a nonlinear eddy viscosity model.

The approaches listed above use statistical inference to construct posterior probability distribution of a quantity of interest based on data. Only calibration of the model coefficients is considered (an L4 uncertainty); typically a simple discrepancy function is defined in the process and often not directly used to make predictions.  
In contrast,  \citet{oliver2009uncertainty}  introduced a Reynolds stress discrepancy tensor
$\bm{\delta}_{\bm{\tau}}$ to account for the uncertainty.  $\bm{\delta}_{\bm{\tau}}$ is a random field described by stochastic differential
equations, which are structurally similar to, but simpler than the Reynolds stress transport equations~\citep[e.g.,][]{launder1975progress}.  They demonstrated preliminary successes of their approach in plane channel flows at various Reynolds numbers. Their framework laid the foundation for many subsequent works in quantifying and reducing RANS model form uncertainties.
 
\citet{dow2011quantification} used full-field DNS velocity data to infer the turbulent viscosity field in a plane channel, based on which they built  Gaussian process models for the discrepancy field. The resulting stochastic turbulent viscosity field was then sampled to make predictions of the velocity field.
\citet{duraisamy2015new,singh2016using} used limited measurements (such as surface pressures, skin friction) to extract 
the  discrepancy field 
and applied it to channel flow, shock-boundary layer interactions, and flows with curvature and separation. 
\citet{xiao2016quantifying} and \citet{wu2016bayesian} used sparse velocity field data to infer the structure of the  Reynolds stress magnitude and anisotropy. 
All these approaches involve large-scale statistical inference
 using adjoint-based algorithms~\citep{giles2003algorithm} or derivative-free, iterative Ensemble Kalman method~\citep{iglesias2013ensemble}.

\begin{figure}
 \centering
 \begin{floatrow}
\subfloat[$C_f$]{\includegraphics[width=5cm]{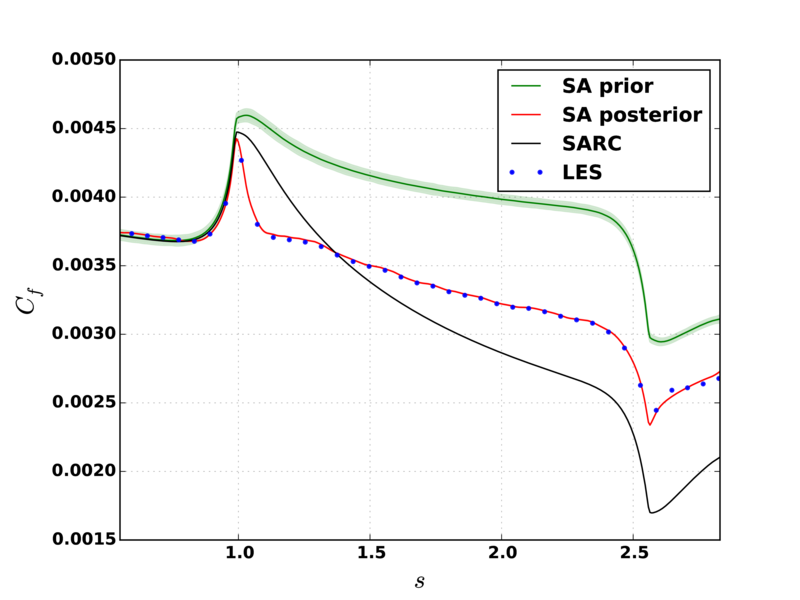}} \hspace{-2.2cm}
\subfloat[$\delta_{MAP}$]{\includegraphics[width=5cm]{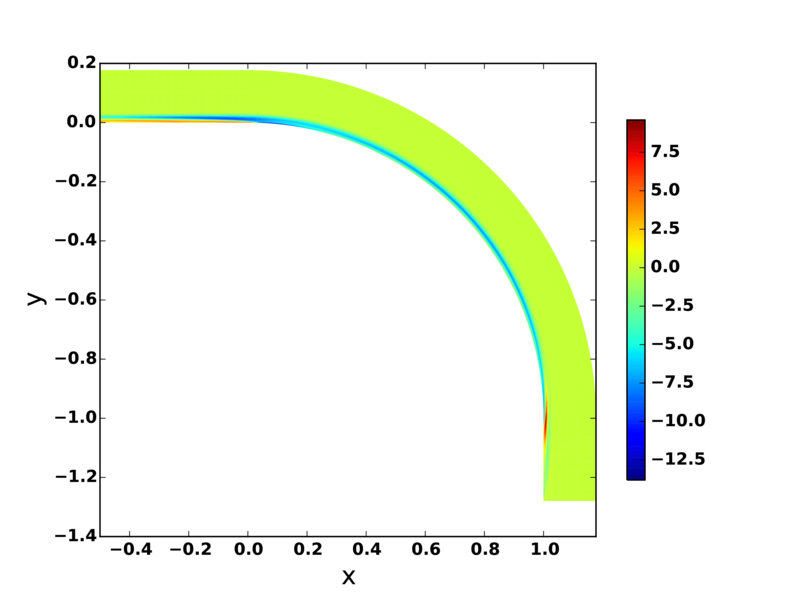}} \hspace{-2.7cm}
\subfloat[$\delta_{SARC}$]{\includegraphics[width=5cm]{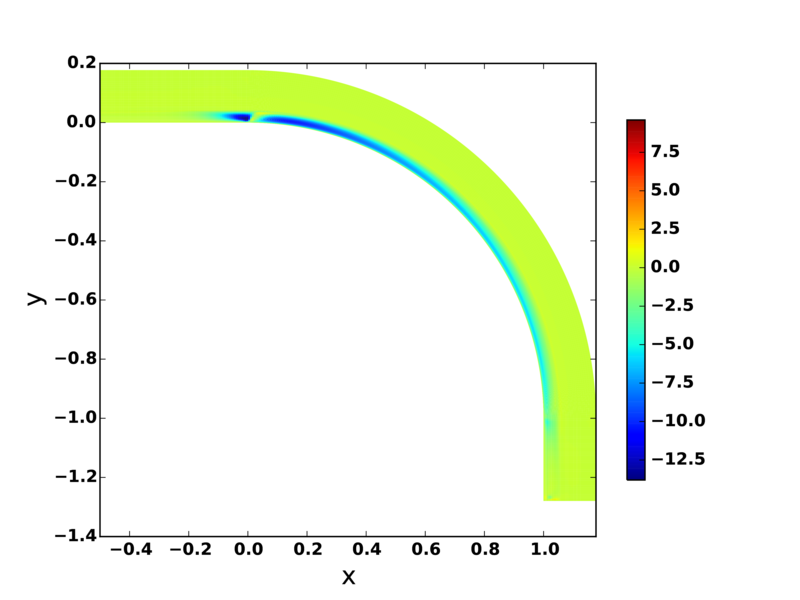}}
\end{floatrow}
\caption{Skin-friction predictions and correction terms for a convex channel.}
\label{fig:convex:sarccomp}
\end{figure}

\begin{figure*}
\centering
\includegraphics[width=0.85\textwidth]{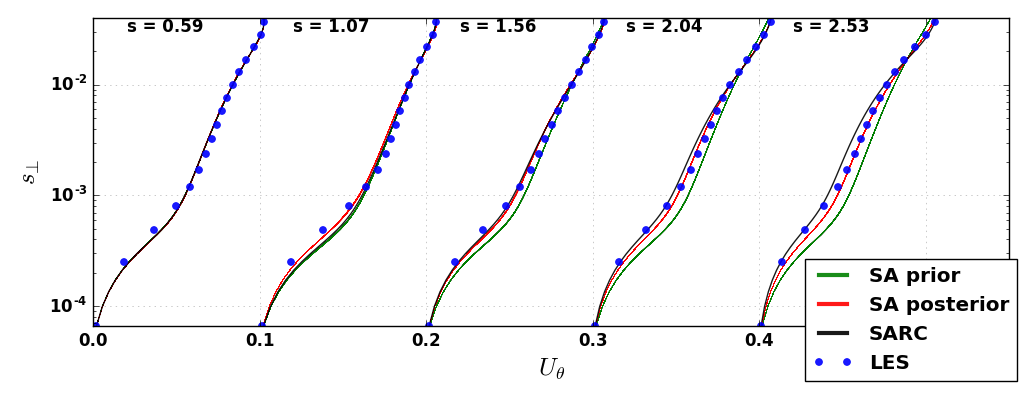}
\caption{Predicted stream--wise velocity at various streamwise $s$ locations  for the convex channel. $s_{\perp}$ refers to the perpendicular distance from the lower wall.}
\label{fig:convex:bl}
\end{figure*}

As a representative example of large-scale inversion, the methodology is illustrated~\citep{singh2016using} for the flow over a curved channel in Figure~\ref{fig:convex:sarccomp}. In this case, the discrepancy function is defined 
 as a multiplier to the production term of the Spalart-Allmaras (SA) turbulence model~\citep{spalart1992one-equation} and extracted by applying statical inversion using the skin friction $C_f$ data obtained via LES on the lower (convex) wall. Figure~\ref{fig:convex:sarccomp} shows the baseline SA model (prior) and posterior (MAP) outputs alongside LES. In
 addition, a  model which includes an analytically-defined correction to the production term of the SA model, namely the SARC model~\citep{shur2000turbulence} is also reported.
Also shown is the inferred correction term, $\bm{\delta}_{MAP}$, and the analytically-defined correction term from the corresponding SARC model.
The trend in $\bm{\delta}_{MAP}$ is consistent with the expectation that the convex curvature reduces the turbulence intensity.
Figure~\ref{fig:convex:bl} shows the variation of the streamwise velocity with respect to the distance from the wall at various streamwise locations. The posterior velocity and Reynolds stresses (not shown) were observed to compare well with the LES counterparts and considerably improved compared to the SA and the SARC prediction, even though only the skin friction data was used in the inference. The results suggest that the SARC model requires improvements in the log layer; a modeler can use the result of the inference to further develop curvature corrections. An alternative viewpoint,
is that this models aims at producing models in which the operators $\mathcal{P}(\mathbf{w})$ used in the (L3) modeling step, i.e. in Eq.~\ref{eq:L3}, are informed by data resulting in~${\mathcal{P}}(\mathbf{w};\bm{\theta})$.

\subsection{Generalizing the embedded discrepancy}
The studies reviewed above inferred the discrepancy  as a spatially varying field using data that are directly relevant to the specific geometry and flow conditions of interest, and therefore  are not easily generalizable. Although efforts such as the scenario averaging \citep{edeling2014bayesian} provide some relief   by incorporating evidence from different datasets,  it is much more desirable to construct discrepancy functions that can be employed within a class of flows sharing similar features (e.g., separation, shock/boundary layer interaction, and jet/boundary layer interaction).
\citet{tracey2013application} used machine learning to reconstruct discrepancies in the anisotropy tensor. Starting with the eigen-decomposition in Eq.~\ref{eq:eigen}, perturbations $\bm{\delta}_{\lambda}$ to the eigenvalues  were derived at every spatial location $\bm{x}$ using a DNS dataset. At this point, $\bm{\delta}_{\lambda}(\bm{x})$ is mapped into a {\it feature} space, $\bm{\eta}(\bm{x})$, consisting
of functions of relevant quantities such as mean velocity gradients and turbulent quantities. The mapping is {\it learned} via Gaussian process regression based on the DNS dataset, and the resulting discrepancy  $\bm{\delta}_{\lambda}(\bm{\eta})$  was found  to
be relatively accurate.
 This work was followed by further developments  in the field of machine learning and lead to a promising research  avenue   that combines turbulence modeling, inference, uncertainty quantification and learning strategies.
 
 \begin{textbox}[h]\section{Machine Learning}
Machine learning is an umbrella term for a wide range of techniques within the broader field of artificial intelligence and, 
 it has been recently rejuvenated by  algorithmic innovations, advances in computer hardware and the enormous growth in the availability of data.

Machine learning can be broadly categorized into unsupervised  and supervised learning. In unsupervised learning, there are no specific targets to predict; the goal is to discover patterns and reduce the complexity of the data. Examples include clustering, i.e. grouping data points based on their  similarity and dimension reduction, i.e. identifying a subset of dependent variables that describe the data. This is in contrast to supervised learning, where the objective is to construct a mapping of the inputs and the outputs. When the output is categorical, supervised machine learning strategies are also referred to as {\it classification} strategies; when the output is continuous these methods are referred to as {\it regression} approaches. The  latter is of particular interest in the context of turbulence modeling. Example techniques commonly used in supervised learning (including both classification and regression) are random forests, support vector machines, and neural networks.

 Neural networks are receiving considerable attention  because of their capability to  approximate complex functions in a flexible form. Typically, an extremely large number of calibration coefficients have to be determined to train a neural network, but efficient algorithms are available for implementation on modern computer architectures. From a mathematical perspective, Neural networks involve the composition of nonlinear functions. 
Starting with an example of a linear model, consider a data set $\bm{\theta}$ and a vector of inputs, or features, $\bm{\eta}$. A linear model for the output $\bm{\delta}(\bm{\eta})$ can be constructed considering
$\bm{\delta}(\bm{\eta}) = \bm{W} \bm{\eta} + \bm{\beta}$, where the weight matrix $\bm{W}$ and the bias vector $ \bm{\beta}$ are obtained by solving an optimization problem that minimizes the overall difference between $\bm{\delta}$ and $\bm{\theta}$. This process is called model training or {\em learning}. However, such a simple model may lack the flexibility to represent a complex functional mapping, and therefore {\em intermediate} variables  (layers) $\bm{\ell}$ are introduced:

\[
\bm{\ell} = \sigma \left( \bm{W}^{(1)}  \bm{\eta} + \bm{\beta}^{(1)}\right) \quad \text{and} \quad   \bm{\delta}(\bm{\eta}) = \bm{W}^{(2)} \bm{\ell} + \bm{\beta}^{(2)},
\]

\noindent where $\sigma$ is a user-specified \emph{activation function} such as the hyperbolic tangent. The two-layer network written as composite function is
\[
\bm{\delta}(\bm{\eta}) = \bm{W}^{(2)} \sigma \left(\bm{W}^{(1)}  \bm{\eta}  + \bm{\beta}^{(1)}\right) +  \bm{\beta}^{(2)}.
\]

 The composition of several intermediate layers results in a {\it deep neural network}, which is capable of efficiently representing arbitrary complex functional forms.

\end{textbox}

\subsection{Modeling using machine learning}

\begin{figure}[!htbp]
  \centering
\includegraphics[width=0.85\textwidth]{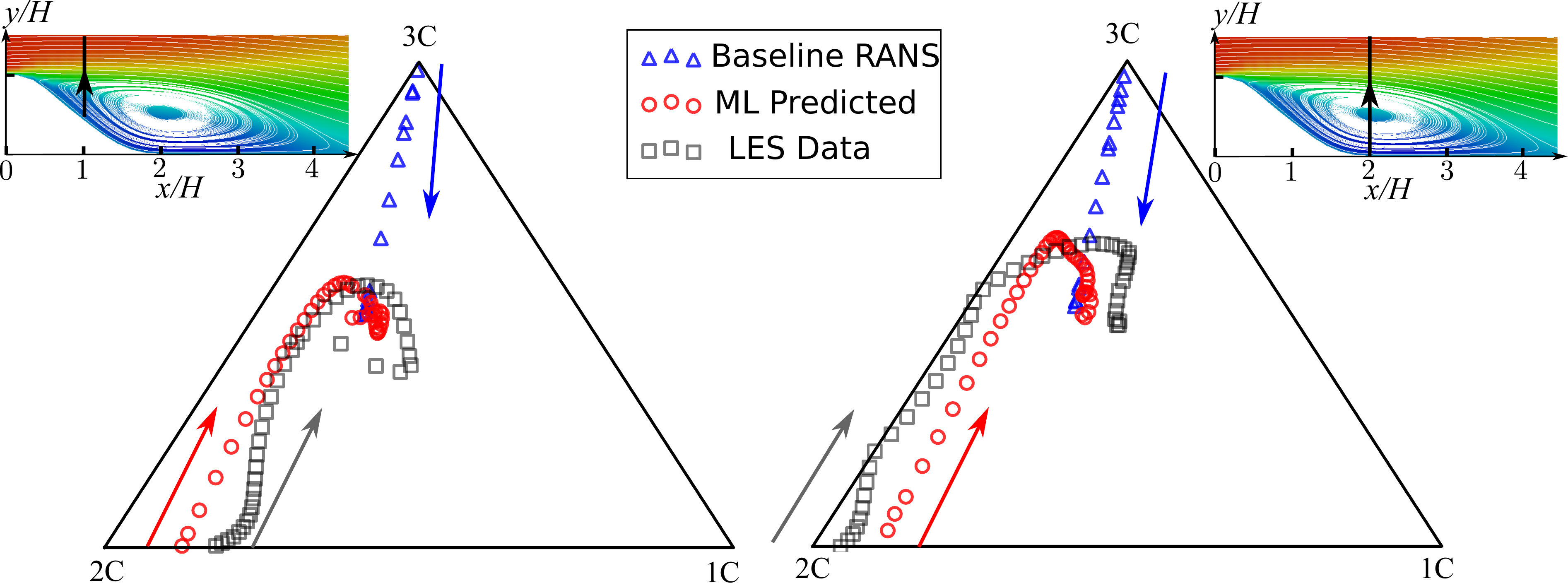}
  \caption{Anisotropy at locations (indicted in the insets) in the flow over periodic hills, predicted by using a random forest model trained on several separated flows in significant different \emph{geometries} and  configurations~\citep{wang2017physics-informed}.}
  \label{fig:piml-results}
\end{figure}

\begin{figure}[!htbp]
  \centering
\includegraphics[width=0.9\textwidth]{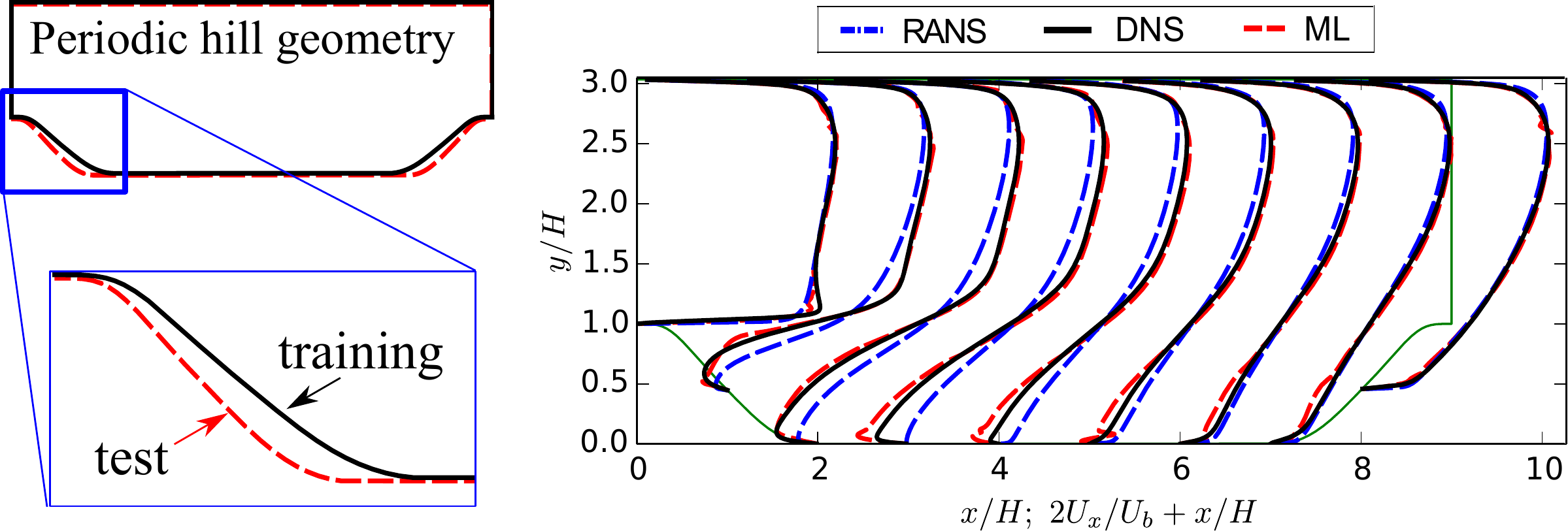}
  \caption{Velocities predicted by using machine-learning corrected Reynolds stresses. The training data is obtained from a flow over periodic hills in a slightly different geometry as shown on the left panel~\citep{wu2018data-driven}.}
  \label{fig:wu-results}
\end{figure}

Machine Learning (ML)  provides effective strategies to construct mapping between large datasets and quantities of interest. ML can be applied directly as a {\it black-box} tool, or in combination with existing models  to provide {\it a posteriori} corrections. 
\citet{tracey2013application} used supervised learning to represent perturbations to Barycentric coordinates. The perturbations are then reconstructed using a machine learning algorithm as a function of local flow variables. While this is an activity at the L3 level (specific to a baseline model), the methodology is generally applicable to any turbulence model. In contrast, \citet{wang2017physics-informed} and \citet{wu2018data-driven} developed a more comprehensive perturbation strategy to predict discrepancies in the magnitude, anisotropy, and orientation of the Reynolds stress tensor. They demonstrated results for two sets of canonical flows, separated flows over periodic hills and secondary flows in a square duct. Representative results are presented in Figures~\ref{fig:piml-results} and~\ref{fig:wu-results}, showing improved predictions of Reynolds stress anisotropy and mean velocities, respectively.

An important aspect of applying machine learning techniques is to ensure the objectivity and 
 the rotational invariance of the learned Reynolds stress models. \citet{tracey2013application},\citet{wang2017physics-informed}, and \citet{wu2018data-driven} used tensor invariants based on the eigen-decomposition of the Reynolds stresses, while for the representation of the stress orientation, both Euler angles and unit quaternions have been considered \citep{wu2017representation}.

As discussed earlier a strategy to develop closures for Reynolds stresses is based on the formulation of a generalized expansion of the Reynolds stress tensor ~\citep{pope75more}. In the assumption that the
stresses only depend on the mean velocity gradient, one can write

 \begin{equation}
\label{eq:astm}
  {\boldsymbol{\tau}} = \sum_{n=1}^{10}c^{(n)}\mathbf{\bm{\mathcal{T}}}^{(n)}
  \end{equation}
  
 \noindent where the coefficients $\bm{c}$ must be obtained from empirical information or further assumptions, while $\bm{\mathcal{T}}$ are known functions of the symmetric and anti-symmetric part of the velocity gradient tensor. In a machine learning framework, one rewrites the expansion as
 
 \begin{equation}
\label{eq:astm}
  \widetilde{\boldsymbol{\tau}} = \sum_{n=1}^{10}c^{(n)}(\bm{\theta},\bm{\eta})\mathbf{\bm{\mathcal{T}}}^{(n)}
  \end{equation}

 \citet{ling2016reynolds} proposed a neural network architecture with embedded invariance properties to learn the coefficients $\bm{c}(\bm{\theta},\bm{\eta})$ with good
 predictive capability but no explicit expression for the resulting model (i.e. any stress evaluation requires the use of the original, calibrated neural network).  In a related effort 
\cite{weatheritt2016novel,weatheritt2017development} used symbolic regression and gene expression programming for defining the coefficients $\bm{c}(\bm{\theta},\bm{\eta})$ in the context of algebraic Reynolds stress models, resulting
in an explicit model form  that can be readily implemented in RANS solvers.  DNS data of flow over a backward-facing step at a low Reynolds number is used for training, while
the flow at a high Reynolds number is predicted. While the results are encouraging, a high level of uncertainty is  observed by applying the resulting model
to the flow over periodic hills.

 The approaches discussed above use the same starting point, an L2 level assumption, and a different set of features ($\bm{\eta}$) and data ($\bm{\theta}$).
 The work of \citet{ling2015evaluation} illustrated a scheme for crafting features based on flow physics and normalizing them  using local  quantities; later work has expanded this approach
 by using invariants of a tensorial set~\citep{ling2016machine}. \cite{wang2017physics-informed} and \citet{wu2018data-driven} used such approaches to predict Reynolds stress discrepancies with a large feature set.
These approaches only use local quantities to construct  the set of features.
 In general, further work  based on modeling non-local, non-equilibrium effects can expand the predictive capabilities of these approaches.
For example, in a traditional eddy viscosity model \citet{hamlington2008reynolds} used variables that account for non-local behavior through streamline integration, which provided an inspiring approach for choosing features in data-driven modeling.
 
\citet{tracey2015machine} performed a proof-of-concept study to learn known turbulence modeling terms from data using neural networks. The neural network based terms were then embedded within an iterative RANS solver and used for predictions, demonstrating the viability of using ML methods in a hybrid PDE/neural networks setting. 
 
\subsection{Combining inference and machine learning}

If machine learning is applied directly to high-fidelity data, inconsistencies may arise between the training environment and the prediction environment. Since turbulence models are typically formulated to accurately represent first and second moments, many latent variables assume an operational rather than a physically precise role in the turbulence models. For instance, the role of the dissipation rate ε in a model is only to provide scale information, and the model is typically calibrated to provide accurate results for the moments, even if the values assumed by ε are different from the true value.

Statistical inference provides a rigorous framework to calibrate models using data while machine learning offers 
a flexible setting to formulate discrepancy functions in terms of a vast number of general features.
The combination of these two strategies yields the promise of deriving effective data-driven closures for turbulence models.   
\citet{duraisamy2015new} and \citet{parish2016paradigm} have explored this avenue.  In the first step, the  spatial structure of the model
discrepancy $\bm{\delta}(\bm{x})$ is extracted using statistical inference from datasets representative of the phenomena of interest.
Machine learning is then used to reconstruct the functional form of the discrepancy in terms of the mean flow and turbulence variables, $\bm{\delta}(\bm{\eta})$.

 \begin{figure*}[!h]
\flushleft
 \begin{floatrow}
\subfloat[]{\includegraphics[width=0.5\textwidth]{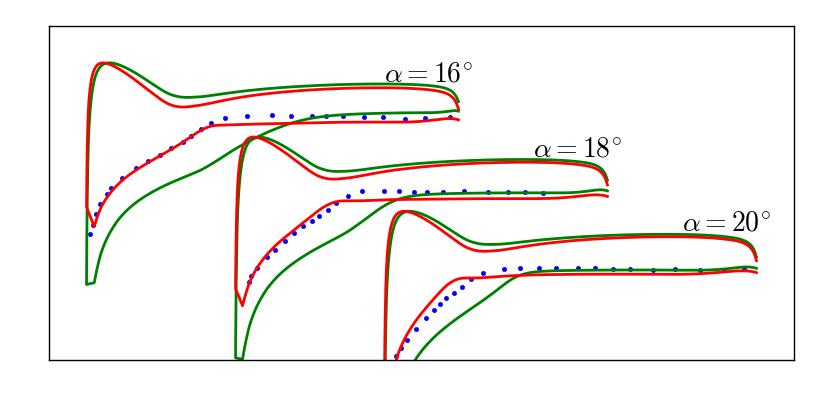}}\hspace{-2 cm}
\subfloat[]{\includegraphics[width=0.3\textwidth,trim={.5cm .5cm .5cm 4.5cm},clip]{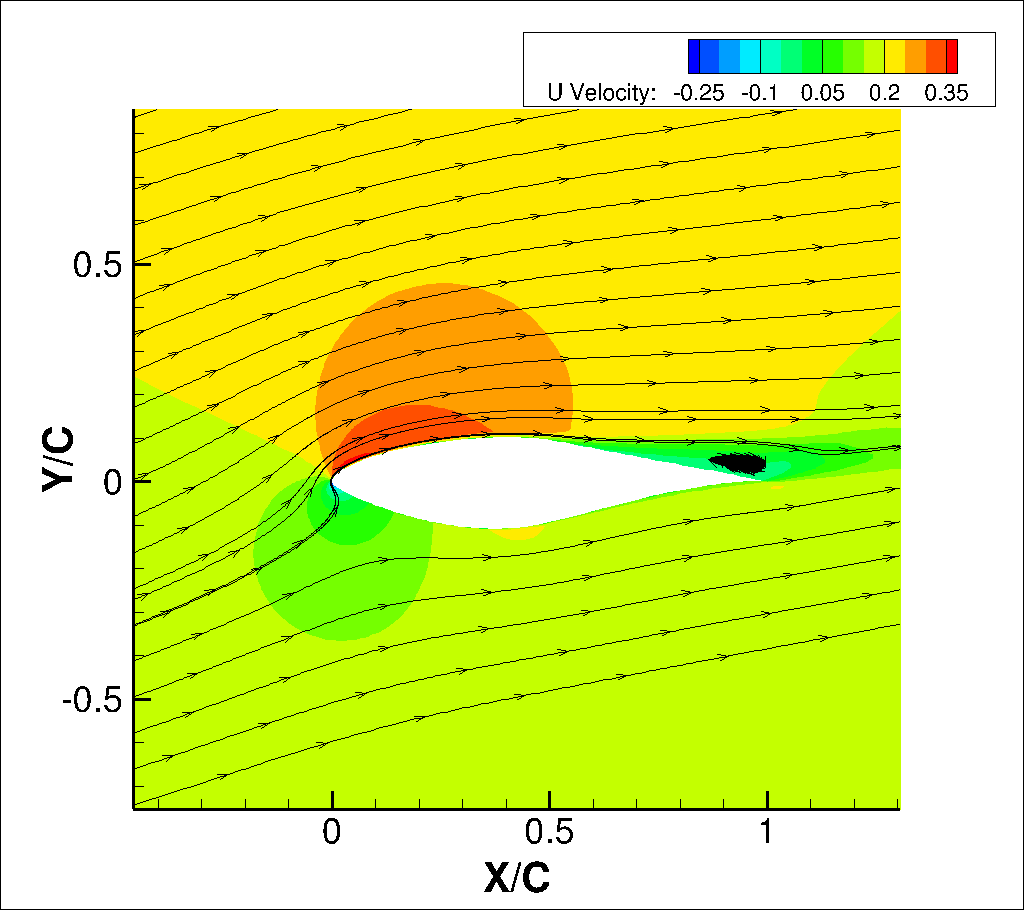}} \hspace{-3cm}
\subfloat[]{\includegraphics[width=0.3\textwidth,trim={.5cm .5cm .5cm 4.5cm},clip]{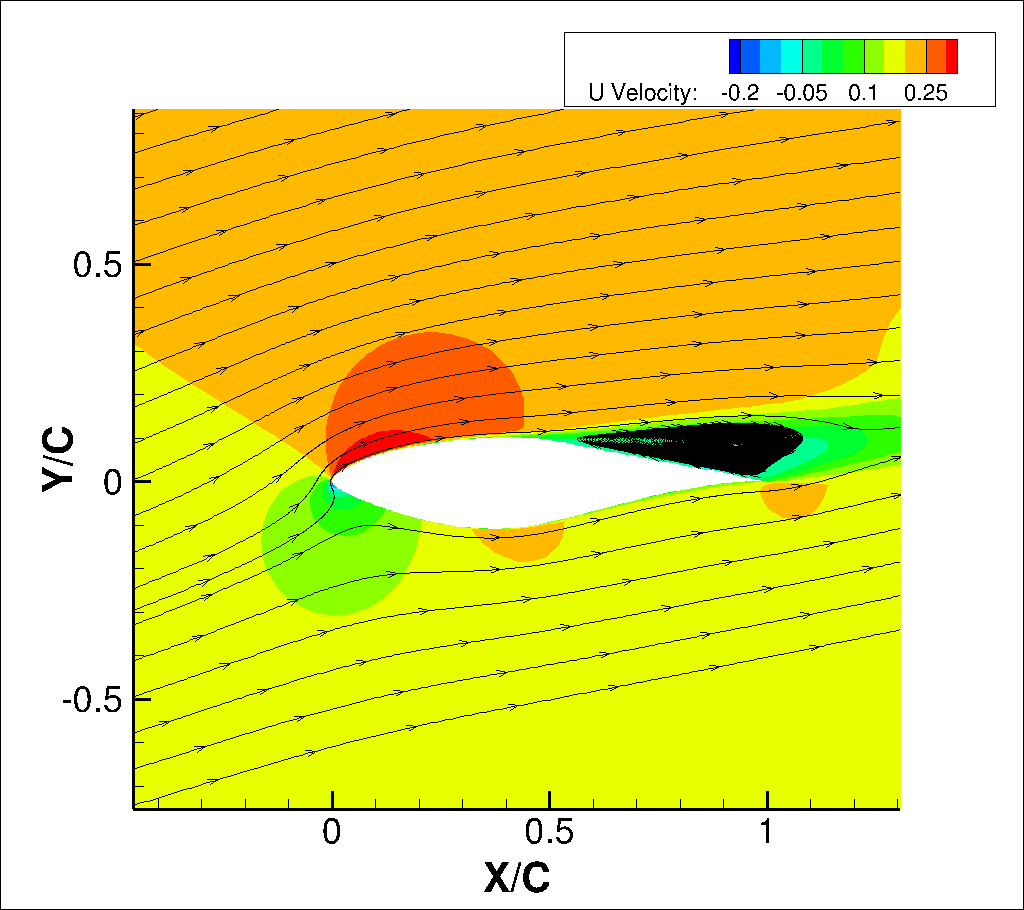}}
 \end{floatrow}
\caption{Example of application of inference and learning to turbulent flows over airfoils~\citep{singh2017machine-learning-augmented}.  (a) Pressure over airfoil surface (Green: Baseline model; Red: ML-augmented  model; Blue: Experimental measurements).(b) Baseline flow prediction (pressure contours and streamlines). (c) Flow prediction using data-driven SA model.}
\label{fig:airfoil}
\end{figure*}

The resulting discrepancy is then embedded in RANS solvers as a correction to traditional turbulence models results in convincing  improvements of the predictions. In \cite{singh2017machine-learning-augmented,singh2017augmentation}, this approach was used for  the simulation of turbulent flow over airfoils. Figure~\ref{fig:airfoil} shows results from a data-driven SA model.
Of particular interest, is that the inversion process does not require extensive datasets, and even very limited experimental measurements, such as the lift coefficient, 
provides useful information that lead to considerable improvements in the predictions with respect to the baseline models.

\section{Challenges and perspectives}

The concurrent enhancements in statistical inference algorithms, machine learning and uncertainty quantification approaches combined with the growth in available data is spurring 
renewed interest in turbulence modeling. We have surveyed efforts in the context of RANS modeling, however data-driven approaches are being pursued in a variety of contexts in fluid dynamics   and for increasingly complex applications, such as multiphase flows. Promising activities in LES include the use on neural networks to model  subgrid-scale stress
 \citep{gamahara2017searching, vollant2014optimal, vollant2017subgrid-scale} and to represent the deconvolution of flow quantities from filtered flow fields  \citep{maulik2017neural}. 
 \citet{ma2015using,ma2016using} used machine learning to model the inter-phase mass and momentum fluxes in multiphase flow simulations.

 In spite of recent successes, several challenges  remain.

\begin{itemize}
\item {\it What data to use?} Databases of experimental measurements and DNS are readily available today, but they might have only limited relevance in specific problems of interest. The quantification of
the {\it information content} in the data is a critical aspect of data-driven models. Ideally, the process of calibration should provide direct indication of the need for additional data or potential overfitting. This is a classic setting to introduce formal design-of-experiments to  drive further data-collection activities.
In addition, the uncertainty present in the data must be accounted for during the inference and learning stages, and eventually propagated to the final prediction to set reasonable expectations in terms of prediction accuracy.
\item {\it How to enforce data-model consistency?} If machine learning is applied directly on a dataset, a compounding problem is the consistency between the data and the models, i.e. the
difference between the learning environment (DNS) and the injection environment (RANS).  It is well known that even if DNS-computed quantities are used to completely replace specific terms in the RANS closure, the overall predictions will remain unsatisfactory \citep{thompson2016methodology, poroseva2016on} due to  the assumptions and approximations at various levels in models, compounded by the potential ill conditioning of the RANS equations~\citep{wu2018on}. Furthermore, scale-providing variables such as the turbulent dissipation rate will be very different in the RANS and DNS context. The addition of the inference step before the learning phase  enforces consistency between the learning and prediction environment.
\item {\it What to learn?} The blind application of learning techniques to predict a quantity of interest based on available data cannot be expected, in general, to produce credible results. A more realistic goal is to focus on learning discrepancy functions and  an appropriate set of features that satisfy physical and mathematical constraints. But how many features are required? And what is the optimal choice for broad application of the resulting calibrated model to different problems? These remain open questions and the subject of on-going investigations.
An alternative, promising approach  is to focus on a specific component  of a closure and introduce correction terms that can be learned from data, as in the example reported earlier corresponding to the curvature correction of the Spalart-Allmaras model ~\citep{singh2016using}. In this case, the learning strategy can provide direct insights to modelers.
\item {\it What is the confidence in the predictions?} Calibrated  models have potentially a limited applicability because of the unavoidable dependency on the data; furthermore, data-driven models might suffer  from  a lack of {\it interpretability}, i.e. the difficulty of explaining   causal relationships between the data, the discrepancy and the corresponding prediction. 
The use of deep learning strategies and vast amount of data in the inference process exacerbates this issue.  
\item {\it What is the right balance between data and models?} Recent works~\citep{schmidt2009distilling,brunton2016discovering,raissi2017physics} have explored the possibility of extracting models purely from data.  It has been  shown that the analytical form of {\em simple} dynamical systems can be  extracted from data and solutions of the Navier--Stokes equations can be reconstructed on a given grid and flow condition. While these methods have focused on {\em rediscovering} known governing equations or {\em reconstructing} known solutions, the possibility of discovering unknown equations and deriving accurate predictive models purely from data remains an open question, though unknown closure terms have been extracted for simple dynamical systems~\citep{pan2018data}. 
 Ultimately, the decision to leverage existing model structures and  incorporate/enforce prior knowledge and physics constraints is a \emph{modeling choice}.
In the limit of \emph{infinite amounts of data}, machine learning could potentially identify universal closure model equations directly from data. On the other hand, relying on machine learning alone, when dealing with large but finite amount of data, problem-specific/spurious laws might be discovered,
resulting in very limited predictive value.
Therefore,  the modeler's choice is dictated by the \emph{relative faith} in the available data and prior model structures, physical constraints, and the purpose of the model itself (e.g. whether the model will be used in reconstruction, or parametric prediction, or true prediction).
 \end{itemize}

In general, a holistic approach that (1) leverages  advances in statistical inference {\em and} learning, (2) combines the data-driven process with the assessment of the {\it information content}, (3) complies with physical and mathematical constraints, (4) acknowledges the assumptions intrinsic
in the closures,  and (5) rigorously quantifies   uncertainties, has the potential to lead to credible and useful models.

In conclusion, we expect a pervasive growth of data-driven models fueled by advances in algorithms and accelerated by novel computer architectures. Moreover, we expect data to profoundly impact
models in  all aspects, i.e.  through parameters $\mathbf{c}$, algebraic or differential operators $\mathcal{P}$ and discrepancy $\bm{\delta}$; this will result in general models written as: 

  \begin{equation}
 \widetilde{ \mathcal M} = \mathcal M (   \mathbf{w}; \mathcal P(  \mathbf{w}; \bm{\theta}  );    \mathbf{c}(\bm{\theta});   \bm{\delta}{(\bm{\theta},\bm{\eta})};   \bm{\epsilon}_{\bm{\theta}}  ).   
 \label{eq:datadrivenfinal} 
 \end{equation}

\noindent and recommend the resulting predictions to be accompanied by explicit uncertainty estimates $\widetilde{\mathbf{q}} = \mathbf{  \widetilde{ \mathcal M}  } + \bm{\epsilon}_{\mathbf{q}}$.

\section*{DISCLOSURE STATEMENT}
The authors are not aware of any affiliations, memberships, funding, or financial holdings that
might be perceived as affecting the objectivity of this review. 

\section*{ACKNOWLEDGMENTS}
GI acknowledges support from the Advanced Simulation and Computing program of the US Department of Energy via the PSAAP II Center at Stanford under contract DE-NA-0002373. KD acknowledges NASA grant NNX15AN98A (tech. monitor: Dr. Gary Coleman) and  NSF  grant 1507928 (tech. monitor: Dr. Ron Joslin). GI and KD acknowledge support from the Defense Advanced Research Projects Agency under the Enabling Quantification of Uncertainty in Physical Systems (EQUiPS) project (tech. monitor: Dr Fariba Fahroo). HX acknowledges  support and mentoring from the Department of Aerospace and Ocean Engineering at Virginia Tech and particularly Prof.\ C.J.~Roy and Prof.\ E.G.~Paterson.
The authors would like to thank Dr.\ J.-L.~Wu, Dr.\ J.-X.~Wang,  Dr.\ J.~Ling, Dr.\ A.~Singh and Dr.\ B.~Tracey for their collaborations and Dr. A. A. Mishra for useful suggestions on the manuscript.

\end{document}